\def\comment#1{}

\newcommand{\ve}[1]{\mathbf{#1}}

\newcommand{\im}{{\rm i}}
\newcommand{\beg}{\begin{eqnarray}}
\newcommand{\eee}{\end{eqnarray}}

\documentclass[prl,letterpaper,twocolumn,aps,epsf]{revtex4}
\usepackage{dcolumn}
\usepackage{amsmath}
\usepackage{graphicx}%
\def\cm#1{}

\begin{document}
\title{Vortex matter and generalizations of dipolar superfluidity concept in layered systems}
\author{Egor Babaev}

\affiliation{
The Royal Institute of Technology, Stockholm, SE-10691 Sweden\\
Centre for Advanced Study, Norwegian Academy of Science and Letters,  N-0271 Oslo, Norway.}

\begin{abstract}
In the first part of this letter we discuss electrodynamics  
of an excitonic condensate in a bilayer.
We show that under certain conditions the system has  a dominant energy scale 
and is described by the effective electrodynamics with  ``planar magnetic charges".
%This suggest that at finite temperature it should possess a magnetic response
%analogues to Halperin-Nelson current response of superconducting films.
In the second part of the paper we point out that  
a vortex liquid state in  bilayer superconductors also possesses dipolar superfluid modes 
and establish equivalence mapping between this state and a dipolar 
excitonic condensate. We point out that
a vortex liquid state in an N-layer superconductor 
possesses  multiple
topologically coupled dipolar superfluid modes
and therefore represents a  generalization of the dipolar superfluidity concept.
\end{abstract}

\maketitle
\newcommand{\la}{\label}
\newcommand{\aaa}{\frac{2 e}{\hbar c}}
\newcommand{\Pfaff}{{\rm\, Pfaff}}
\newcommand{\kA}{{\tilde A}}
\newcommand{\G}{{\cal G}}
\newcommand{\cP}{{\cal P}}
\newcommand{\M}{{\cal M}}
\newcommand{\E}{{\cal E}}
\newcommand{\btd}{{\bigtriangledown}}
\newcommand{\W}{{\cal W}}
\newcommand{\X}{{\cal X}}
\renewcommand{\O}{{\cal O}}
\renewcommand{\d}{{\rm\, d}}
\newcommand{\bfi}{{\bf i}}
\newcommand{\e}{{\rm\, e}}
\newcommand{\bfx}{{\bf \vec x}}
\newcommand{\bfn}{{ \vec{\bf  n}}}
\newcommand{\bfs}{{\vec{\bf s}}}
\newcommand{\bfE}{{\bf \vec E}}
\newcommand{\bfB}{{\bf \vec B}}
\newcommand{\bfv}{{\bf \vec v}}
\newcommand{\bfU}{{\bf \vec U}}
\newcommand{\bfp}{{\bf \vec p}}
\newcommand{\f}{\frac}
\newcommand{\bxy}{{\bf B}_{in}}
\newcommand{\bfA}{{\bf \vec A}}
\newcommand{\non}{\nonumber}
\newcommand{\be}{\begin{equation}}
\newcommand{\ee}{\end{equation}}
\newcommand{\ba}{\begin{eqnarray}}
\newcommand{\ea}{\end{eqnarry}}
\newcommand{\bastar}{\begin{eqnarray*}}
\newcommand{\eastar}{\end{eqnarray*}}
\newcommand{\half}{{1 \over 2}}

The progress in semiconductor technology has
made it feasible to produce bilayers
where the interparticle distance is larger than the separation of the layers and 
which are expected to have interlayer excitonic states \cite{bilayers1a,shev,josephson,ABPRL,bilayers2-1} (as shown in Fig. 1) 
with a significant life time.  Some signatures of  Bose-Einstein condensate of such excitons
were reported  \cite{UCSD}. 
%This subject has also been attracting 
A renewed theoretical interest in these systems \cite{bilayers1a,shev,josephson,ABPRL,bilayers2-1} is
focused on identification of possible unique properties of such condensates 
which can set them apart from other families of quantum fluids.
In \cite{ABPRL} it was discussed in great detail that the spatial separation of the positive and  negative charges,
in an exciton in a bilayer, makes the phase $\theta$ of the condensate transform
as $\nabla \theta \rightarrow \nabla \theta-  e {\bf A}({\bf r_+}) + e {\bf A}_-({\bf r_-}) $.
The dipolar excitonic condensate (DEC) therefore features a coupling to a {difference} of vector potential values 
at positions in different layers ${\bf r_+}$ and ${\bf r_-}$ \cite{bilayers1a,shev,ABPRL}. In the case of a small layer separation, $d$,
one finds ${\bf A}({\bf r_+}) - {\bf A}_-({\bf r_-})  \approx d \partial_z  {\bf A}({\bf r})$.
Therefore a static in-plane magnetic field ${\bf H}^{ext}$  produces
excitonic currents and
the system is described by the following
free energy density \cite{bilayers1a,shev,ABPRL}:
\beg
{\cal F} =
\f{\rho}{2} \left(\nabla\theta + ed [{\hat{R}}{\bf H}^{ext}]\right)^2 
\label{model}
\eee
where the operator ${\hat{R}}$ rotates a vector 90 degrees counterclockwise:
${\hat{R}}(a,b,c)=(-b,a,c)$, $\rho$ is  the phase stiffness, $e$ is the electric charge and $d$
is the separation of the layers. 
A particularly interesting aspect of this observation is that although an exciton
is, by definition, an electrically neutral object, it has the dipolar coupling to a gauge field
and its effective low-energy model has a symmetry different from the so-called
``global" $U(1)$ symmetry which one finds in ordinary neutral superfluids. It is
also different from the ``gauged" $U(1)$ symmetry which one finds in superconductors.
%The unusual form of the model suggest considerable possibilities for emerging new physics.
%The goal of this paper is to examine how this type of gauge field coupling manifests itself in the
%physics of topological excitations and 
%in the magnetic response of the system in the
%presence of thermal fluctuations.
\begin{center}
\begin{figure}[htb]
\centerline{\scalebox{0.4}{{\includegraphics{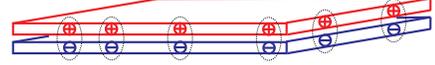}}}} 
\caption{
(color online) A schematic picture of DEC.
An exciton forms as a result of the pairing of an electron in an electron-rich layer and a hole
 in a hole-rich layer. }
\end{figure}
\end{center}
%In the first part of this paper we focus on .
To describe topological defects 
we should include dynamics of the gauge field in the model (\ref{model}).
%An appropriate
%starting point to model the problem is 
Consider
a system of two thin parallel layers with positive and negative charges bound in pairs
(as shown in Fig. 1). 
%The magnetic field
%configuration %is three-dimensional and 
%obeys Maxwell equations outside the layers. 
Here,
\beg
&&\nabla \times {\bf B}={\bf J}=
e[\delta (z_-) -\delta(z_+)]  \times \nonumber \\
&&\Bigl\{
\f{i}{2}
(\psi\nabla\psi^*-\psi^*\nabla\psi)- 
e|\psi|^2[{\bf A}({\bf r}_+)-{\bf A}({\bf r}_-) ]
\Bigr\} %- \nonumber \\
%&&e\delta (z_+)
%\Bigl\{
%\f{i}{2}
%(\psi\nabla\psi^*-\psi^*\nabla\psi)
%-e|\psi|^2({\bf A}({\bf r}_+)-{\bf A}({\bf r}_-) )
%\Bigr\} \nonumber
\eee
 where ${\bf J}$ is the electric current, $\psi$ is the DEC
order parameter, $z_{(+,-)}$ are the $z$-axis positions of the upper and lower planes
and  ${\bf r}_{(+,-)}=(x,y,z_{(+,-)})$. Topological defects in this
system correspond to the situation when the phase, $\theta$, 
of the  order parameter, $\psi$, changes by $2\pi n$. 
A configuration of accompanying magnetic field
should be determined by minimization
of the energy taking into account (i) the  kinetic energy of  currents in two planes, (ii) the
potential energy of the corresponding Ginzburg-Landau functional,
(iii) the energy of the {\it three dimensional} magnetic field configuration.
This problem is non-local
and the field-inducing current itself depends  on the gauge field,
i.e., should be determined 
in a  self-consistent way.
However we show here that there is a regime when 
the system is  accurately described by an
unusual, on the other hand tractable effective model.
That is, let us consider the situation where
  the separation of the layers $d$ is small compared to the system size
and ${\bf A}({\bf r_+}) - {\bf A}_-({\bf r_-})  \approx d \partial_z  {\bf A}({\bf r})$.
% and $d \ll (\rho e^2)^{-1}$.
Then in the hydrodynamic limit the effective model is
\beg
 {\cal F}_{eff} =\frac{1}{2}
%\left\{
\delta(z) {\rho} \left(\nabla\theta + ed [{\hat{R}}{\bf B}_{in}({\bf r})] \right)^2 + \frac{1}{2}{\bf B}^2({\bf r}) 
%\right\}
\label{model}
\eee
Where  ${\bf B}_{in}$ is the field in the dipolar layer.
Consider a vortex with  $\Delta\theta=2\pi$. Then 
$\nabla\theta=\f{1}{r}{\bf e}_\theta$, which
% where $\theta=\tan^{-1} (y/x)$.
 produces
a logarithmic divergence of the energy in a neutral system. 
%However %In contrast to a neutral $U(1)$ superfluid,
%eq. (\ref{model}) has an
%extra degree of freedom associated
%with the gauge field.
  However eq. (\ref{model}) suggests that for a given  phase winding, the
system can minimize its energy by  generating 
a certain configuration of  ${\bf B}_{in}$. %, in the dipolar layer.
Note that the second term in (\ref{model})  depends quadratically on
 ${\bf B}$ and  does not 
allow  a configuration of $\bxy$ which would completely compensate the phase gradient
in the first term.
%This is in contrast to a vortex in a superconducting film
%where the gauge field  compensates the phase gradient at a certain distance
%making the vortex  a finite-energy object.
A configuration of the interplane field, 
% ${\bf B}_{in}$ 
 which would partially compensates 
the divergence
caused by $\nabla\theta$, 
should satisfy the condition:  
%\beg
$ed ({\hat{R}}\bxy) \propto \f{1}{r} {\bf e}_\theta \ (r\gg r_{core}).$
%\eee
This implies the following self-induced interplane field:
%should have the following geometry: 
\beg
\bxy = \alpha \f{1}{ed}\f{{\bf r}}{r^2}  \ \ \ \ \ \  [\alpha<0, {\bf r}= (x,y,0)]
\label{mfield}
\eee
At a first glance, eq. (\ref{mfield}) appears to violate
 the condition that the magnetic field should be divergenceless. % in the absence of magnetic monopoles.
This problem is resolved by including in the picture the ``out-of-plane" magnetic
fields. These, as schematically shown in Fig. 2, can  restore the  $\nabla \cdot  {\bf B}=0$
constraint in three dimensional physical space while permitting the in-plane
field to have the form required by (\ref{mfield}) with a natural cutoff close to the vortex core.
%Thus the  magnetic field in the dipolar bilayer 
%can in principle have the form given in (\ref{mfield}).
%Let us now find the energetically preferred solution.
From the $\nabla \cdot  {\bf B}=0$ 
condition % which  the magnetic field should obey in three dimensions,
it follows that both the interlayer ${\bf B}_{in}$ field
and the out-of-plane field carry the same flux. However
for a given magnetic flux, the
out-of-plane field has the freedom to spread in the 
positive and negative $z$-axis directions.  
Since the magnetic flux is $\int {\bf B} \cdot d{\bf S}$,
while the magnetic field energy is $(1/2)\int {\bf B}^2$,
the out-of plane field
%having the freedom  to spread along $z$-axis
 will have a finite value of the integral
$(1/2)\int_{{\bf r}, z \notin [z_-..z_+] } {\bf B}^2$ over entire three dimensional 
space excluding the bilayer space.
On the other hand the  magnetic field
inside the bilayer behaves as  $ |{\bf B}_{in}| \propto 1/r$ and thus has 
logarithmically divergent energy 
 $(1/2)\int_{{\bf r}, z \in [z_-..z_+] }  {\bf B}^2$.
Therefore from the
condition $\nabla \cdot  {\bf B}=0$ and geometry of the problem
it follows that  the energy of out-of-plane 
magnetic fields is negligible compared to  the energy 
of magnetic field in the dipolar bilayer
for a sample with $d$  small compared to the system size. % and $d \ll (\rho e^2)^{-1}$.

Thus we have identified the regime where the dynamics
of the magnetic field is dominated by its most energetically costly (weakly divergent) interlayer
part,  ${\bf B}_{in}$, which leads to an interesting 
two-dimensional effective model where the magnetic
field ${\bf B}_{in}$ is {\it not}  subject to
the constraint that $\nabla \cdot {\bf B}_{in}=0$: 
\begin{center}
\begin{figure}[htb]
\centerline{\scalebox{0.37}{{\includegraphics{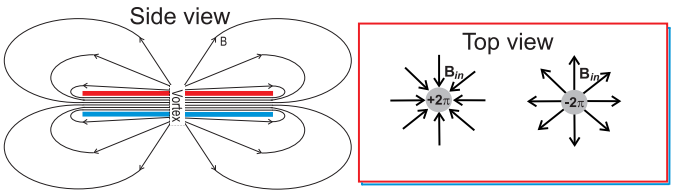}}}} 
\caption{ (color online) 
A schematic picture of a magnetic field configuration in DEC with  vortices.
Side view: the out-of-plane magnetic  field  ${\bf B}$ carries the same flux as the interplane
field ${\bf B}_{in}$, however it has the freedom to minimize energy by  spreading above and below the top and bottom layers.
Top view: the in-plane magnetic field ${\bf B}_{in}$ configuration for vortices
with phase windings $\Delta \theta = \pm 2\pi$
%As explained in the text one can neglect the energy
%of the out-of-plane field  compared to the energy of the field in the dipolar bilayer.
}
\end{figure}
\end{center}
\beg
{\cal F}_{eff}(x,y) = \f{\rho}{2} \left(\nabla\theta + ed [{\hat{R}}{\bf B}_{in}(x,y)] \right)^2 + \f{d}{2} {\bf B}_{in}^2(x,y) 
\label{modeleff}
\eee
Let us consider a vortex with a phase winding $\Delta \theta=2 \pi n$ in the model (\ref{modeleff}).
The coefficient, $\alpha_{min}$,  for the ansatz (\ref{mfield}) which
minimizes the spatially integrated free energy density (\ref{modeleff})
for vortex with $\Delta \theta=2 \pi n$ is
%\beg
$\alpha_{min}^{\Delta \theta=2 \pi n}=-{n \rho}[{\rho+{(e^2d)^{-1}}}]^{-1}$.
%\eee
Thus the vortex % with a phase winding $\Delta \theta=2 \pi n$
has the following configuration of the %self-induced 
magnetic field
\beg
\bxy^{\Delta \theta=2 \pi n} = -n\frac{\rho}{\rho+(e^2d)^{-1}}\f{1}{ed}\f{{\bf r}}{r^2}  
\eee
where ${\bf r}=(x,y,0)$ (shown on  Fig. 2). 
%The configuration of the magnetic field for vortices with  $\Delta \theta=\pm 2 \pi$ is
%illustrated in Fig. 3.
Therefore, these defects
emit a quantized {\it radial} magnetic flux
$\Phi= |{\bf B}({\bf r})| 2\pi r d$. The quantization condition is:
$
\Phi^{\Delta\theta=2\pi n}=- \f{2\pi}{e}  \frac{\rho }{\rho+(e^2d)^{-1}}n.
$
Thus a ``dipolar flux quantum" is
\beg
\Phi_d^0=\f{2\pi}{e}  \frac{\rho e^2d}{\rho e^2d+1} < \Phi_0,
\label{dquant}
\eee
where $\Phi_0=2\pi/e$ is the standard magnetic flux quantum.
In  a way these topological excitations play a role
of positive and negative ``magnetic charges"
in the effective  planar electrodynamics of the model (\ref{modeleff}).

The energy of such a vortex  placed
in the center of a circle-shaped system of a radius $R$ is:
\beg
{ E} 
\approx 
%\Biggr[
%\left(1 - \f{\rho e^2d}{\rho e^2d +1} \right)^2
\f{\pi \rho n^2}{1+ \rho e^2d} 
\ln\f{R}{r_{core}}
\label{vortenerg}
\eee
where  $r_{core}$ is the vortex core size. 
Even though this energy  is logarithmically
divergent, these vortices can in principle be induced in a DEC by
an external in-plane magnetic field because  their magnetic field also has
divergent energy. This provides a possibility to obtain a 
negative contribution to the Gibbs energy from $\int_{\bf r} {\bf B}\cdot{\bf H}$. For example, in principle, an
experimentally realizable configuration
of induction elements can produce 
the following external  in-plane field
$ {\bf H} \approx \gamma \frac{{\bf r}}{r^2}; \ \ \ [{\bf r}=(x,y,0); \ r > r_0] $
%\label{extfield}
%\eee
where $r_0$  is some cutoff length. % which is specific to the geometry of the  induction elements.
Then, for a vortex with $\Delta\theta = 2\pi$, the integral  $\int_{\bf r} {\bf B}\cdot{\bf H}$  provides a negative
logarithmically divergent contribution to the Gibbs energy $G=\int_{\bf r} ( {\cal E}-{\bf B}\cdot{\bf H})$ .
There is a critical value of the coefficient, $\gamma$ 
at which the creation of a vortex becomes energetically favorable 
(below we set $r_0 \approx r_{core}$ without loss of generality).
In the regime  $\rho e^2d \ll 1$ we obtain $\gamma_c \approx \f{1}{2ed}$.
In general case the  energetic favorability of a vortex state depends on the integral $\int_{\bf r}{\bf B}\cdot{\bf H}$.
An external field which is significantly stronger  
than that corresponding to $\gamma_c$
favors a higher phase winding number of
the induced vortex structure, however on the other hand,
the vortex energy depends quadratically on $n$ [see eq. (\ref{vortenerg})].
Therefore, a stronger field should normally produce a state with multiple 
one-dipolar-quantum vortices.

In a planar $U(1)$-symmetric
system, thermal fluctuations can excite finite-energy pairs of vortices with opposite windings. %\cite{KT}. 
In the model (\ref{modeleff})
the interaction between a vortex with $\Delta \theta=2\pi$ located at  ${\bf r}_1$ and an antivortex
with $\Delta \theta =- 2\pi$ located at   ${\bf r}_2$ originates from two sources.  The usual current-current interaction 
produces the attractive force
\beg {\bf F}_{{J}} 
=-2\pi \rho
\left(1 - \f{\rho e^2d}{\rho e^2d+1} \right)^2
\f{({\bf r}_2-{\bf r}_1 )}{|{\bf r}_2-{\bf r}_1 |^2},
\eee 
the other contribution comes
from the ${\bf B}^2_{in}$ term:
 \beg
{\bf F}_{{ B}} 
=-2\pi \rho
%\left[
\f{\rho e^2d}{ [\rho e^2d +1]^2 }
%\right]
\f{({\bf r}_2-{\bf r}_1 )}{|{\bf r}_2-{\bf r}_1 |^2}.
\eee

A system of these vortices and antivortices can be mapped onto a Coulomb gas which at a temperature $T_{KT}$ undergoes a KT transition:
\beg
T_{\rm KT}=
\f{\pi}{2}\rho(T_{\rm KT}) 
%(T_{\rm KT}) 
%\Biggr[
%\left(
\f{1}{1+\rho(T_{\rm KT})
%(T_{\rm KT})
e^2d}
\eee
%This is a monotonically decreasing function
%of the dimensionless parameter $\rho e^2 d$
%which characterizes the strength of the direct coupling to the magnetic field. 
Therefore,  the temperature
of the condensation transition in DEC will be 
suppressed compared to the value of
the condensation temperature
in a neutral system with similar density. 
While the suppression for realistic DECs is tiny, there is an interesting aspect in it
because in contrast to the superfluid density jump $\rho/T_{\rm KT}=2/\pi$ 
in regular superfluids here one should define a ``generalized superfluid density jump" 
which depends on the electric charge and on a nonuniversal paramter: the layer separation.

As is well known,
one of the ways to detect a KT transition/crossover in 
superconducting films is associated with a peculiar
reaction to an  applied current \cite{Halperin}. 
In a superconducting film 
 an external current results in Lorentz forces
acting on a vortex and antivortex
in opposite ways, causing a pairbreaking effect.
Free vortices create  dissipation which is manifested
in $IV$ characteristics \cite{Halperin}. In DEC,
vortices have magnetic field which can be viewed as that of ``magnetic charges"  and, correspondingly,
they are  sensitive to  an applied external
uniform in-plane magnetic field.  Such a field, at finite temperature, should
create a KT-specific modification
of the zero-temperature response discussed in \cite{ABPRL}.
%and, therefore, provides a probe of the
%specific KT transition outlined above. 

We note that because of small carrier density and layer
separation in the presently available semiconductor bilayers \cite{UCSD} 
 a dipolar vortex would carry only about 10$^{-7}$ flux quanta, which, though 
may be  resolved with a modern SQUID, makes observation
of these effects difficult. This raises the question if there could be strong-coupling
dipolar superfluids in principle.
Below  we show that the concept to dipolar superfluidity arises in a system
principally different from DEC, %and where the dipolar coupling can be much stronger.
%where it has a different origin, 
without interlayer pairing problem (which limits
dipolar coupling strength in DEC).  This provides
a possibility to have larger 
%effective dipolar coupling strength 
%and thus larger 
flux of a dipolar vortex and more pronounced phenomena associated with it. Moreover there the concept of dipolar superfluidity 
allows  generalization. 

Consider a layered superconductor (LSC), i.e. a multiple superconducting layers separated by insulating layers (to effectively eliminate
interlayer Josephson coupling) so that the layers
are only coupled by the gauge field. This system has been extensively studied in the past \cite{RMP}. 
In the hydrodynamic limit its free energy density is
\beg
\label{gl_action}
F&=& \sum_{i=1}^N \frac{1}{2}\delta(z_i){\Big|\Big(\nabla -\im e\ve{A}(x,y,z_i)\Big) \psi_i (x,y,z_i)\Big|^2}
\nonumber \\ &+& \frac{(\nabla\times\ve{A}(x,y,z))^2}{2}%\\
\eee
where $\psi_i(x,y,z_i)=|\psi_i(x,y,z_i)|e^{i\theta_i(x,y,z_i)}, z_i+d_i=z_{i+1}.$
%\ \Psi^2 =\sum_i |\psi_i|^2$.
This model indeed does not feature dipolar superfluidity of the DEC type. Here, the main
distinction is the reaction to the external field which is  screened because of the Meissner effect
(the effective screening length in a planar superconductor is $\lambda^2$/[layer thickness]). 
However there are  situations where the superconductivity  can be
eliminated  in this system  by  topological defects. 
That is, if to apply an external field along $z$-direction 
one can produce a lattice of  vortex lines with phase winding in each layer
 ($\Delta\theta_1(z_1)=2\pi,...,\Delta\theta_N(z_N)=2\pi$) %theading the system in $z$-direction
 \cite{RMP}.
Enclosed magnetic flux gives such a vortex line a finite tension.
 At elevated temperatures the lattice of these vortices melts but importantly
normally there is a range of parameters (which depends on the strength of the applied field
and temperature) where the vortex lines forming a liquid retain tension \cite{sudbo,RMP}.
If the vortex lattice is not pinned or if one has a tensionfull vortex liquid the charge transfer
in superconducting layers is dissipative. However in such situations
 a system nonetheless retains broken symmetries associated with the phase differences between the order parameters, and correspondingly dissipationless countercurrents \cite{myworks},  in this particular case a broken symmetry is retained in the phase difference between the layers.
Physically the situation which occurs is the following: consider the  $N=2, |\psi_1|=|\psi_2|=|\psi|$ case.
Currents in individual layers move unpinned vortices 
which  produce  dissipation. However as long as a vortex line threading the system has a finite tension, equal countercurrents
in different layers deform but not move a vortex line and therefore do not create dissipation.
For small layer separation  $d$,  the dissipationless counterflows can
approximately be described by extracting phase difference terms \cite{myworks}.
Then the part of the model (\ref{gl_action})
which retains a broken symmery is:
\begin{eqnarray}
F_d &\approx& \f{1}{4} |\psi|^2 \Big[ \nabla (\theta_1(x,y,z_1)- \theta_2(x,y,z_2)) \nonumber \\ & &
- e  ({\bf A}(x,y,z_1)-{\bf A}(x,y,z_2)) \Big]^2
 + \frac{d}{2}\ve{B}_{in}^2.
\label{dipolarSC}
\end{eqnarray}
The vortices in  this system are related to dipolar vortices in DEC.  
%First, consider the  case $N=2, |\psi_1|=|\psi_2|=|\psi|$.
% and the layer  separation $d$. 
The simplest vortices with a topological charge in the phase difference are ($\Delta\theta_1=\pm 2\pi,\Delta\theta_2=0$)
 and ($\Delta\theta_1=0,\Delta\theta_2=\pm 2\pi$). We denote them by a pair of integers $(\pm 1,0)$ and $(0,\pm 1)$. For a vortex 
$(\pm 1,0)$ the currents in two layers are: ${\bf j}_1=e|\psi|^2\nabla\theta_1-e^2|\psi|^2{\bf A}(x,y,z_1); \ {\bf j}_2=-e^2|\psi|^2{\bf A}(x,y,z_2)$.
Like in a DEC, in the case of a finite $d$, the vortex features in-plane radial magnetic field $[{\hat R}{\bf B}_{in}] \approx \partial_z {\bf A}({\bf r})$. 
These vortices can be induced by an external in-plane magnetic field, like in a DEC. However there are indeed also principal differences.
LSC is a system with more degrees of freedom and the above considerations apply only to the state
when superconductivity in individual layers  is removed 
e.g. by a molten lattice of (1,1) vortices, or thermally excited $(\pm 1, \pm 1)$ vortices. 
Note also that the molten lattice of (1,1) vortices does not automatically preclude a formation of an ordered state
of $(\pm 1,0)$ and $(0,\pm 1)$ vortices because  a vortex (1,1)
does not have a topological charge in the phase difference sector $\nabla(\theta_1 -\theta_2)$,  and their disordered states can not eliminate corresponding phase stiffness.
The density of (1,1) vortices and
correspondingly the temperature of their lattice melting is controlled by the strength of the magnetic field in $z$-direction
while the density of $(\pm 1,0)$ and $(0,\pm 1)$ vortices is controlled by the in-plane magnetic field.
The system therefore possesses a control parameter which allows for the
ordered structures of $(\pm 1,0)$ and $(0,\pm 1)$ vortices to coexist with a liquid  of $(1,1)$ vortices. 

 The dipolar superfluidity in LSC has a number of  detectable physical consequences. Namely the 
LSC in the vortex liquid state should have a dipolar
superfluid response, analogous to that discussed in great detail in \cite{ABPRL} . Also the 
% vortices $(\pm 1,0)$ and $(0,\pm 1)$ possess features of the  ``planar magnetic charges" and correspondingly 
 system in the vortex liquid state should possess aspects
of planar electrodynamics with effective magnetic charges discussed above in connection with DEC
which may be observable in the temperature dependency of the dipolar response.
The general case of N layers, especially with the variable interlayer distances and  the condensate densities,
has much richer structure than DEC because the dipolar superfluid modes are multiple and 
coupled. That is, for N layers one can have multiple combinations of counterflows in different layers which will not
move a tensionfull vortex line, therefore one has to consider all possible combinations of phase differences to describe
dipolar modes. In the limit of very small $d$ the kinetic terms for countercurrents approximately can be expressed as:
\begin{eqnarray}
F_d^{CF} &\approx& \sum_{i,j=1}^N \f{|\psi|^2}{{4 N}}  
 \Big[ \nabla (\theta_i(x,y,z_i)- \theta_j(x,y,z_j)) \nonumber \\ & &
- e  ({\bf A}(x,y,z_i)-{\bf A}(x,y,z_j)) \Big]^2.
\label{dipolarSC}
\end{eqnarray}
%Here $\Psi^2 =\sum_i |\psi_i|^2$.
 This generalization of DEC may be called ``multiflavor dipolar condensate". 

In the both cases of DEC and LSC the dipolar superfluidity will be destroyed by interlayer tunneling, but for small interlayer 
tunneling some of its physical manifestations will in some cases remain \cite{josephson}.

In conclusion we have  considered  the possible physical effects 
dictated by the symmetry and dynamics of a gauge field
in  dipolar condensates. 
For DEC, we started  by constructing 
an effective model, based on symmetry and energy scales of the problem. 
In this  framework 
 we described topological defects
which emit a nonuniversally quantized {radial} magnetic flux.
We pointed out that because of the existence of well separated energy scales
in the regime, when interlayer distance is smaller than
other length scales in the problem,
the system  possesses  effective 
planar electrodynamics with  ``magnetic
charges" (planar analogue of magnetic monopoles) 
arising from topological excitations. Therefore at finite temperature an applied  inplane magnetic
field should have a vortex-antivortex pairbreaking effect.
Experimentally it may manifest itself through a counterpart  of Halperin-Nelson
response which may be observable in systems with sufficiently strong dipolar coupling.
In the second part of the paper we map DEC onto the tensionful vortex liquid state in 
layered superconductors. In that system
we point out the emergence of a dipolar superfluidity
which has a different origin because
there is no interlayer pairing and carriers in layers have the same sign of electric charge.
There the analogue of the
dipolar superfluidity arises as a consequence of the fact that a molten 
lattice of composite vortices makes currents in individual layers dissipative while the counter-currents
in different layers remain dissipationless. In these systems there is no  need for interlayer pairing and carrier  density is higher
so the effective dipolar coupling may be relatively large. 
Besides that, the dipolar
response can be used as an experimental tool to study vortex liquids in LSC, e.g. to unequivocally distinguish a tensionfull
vortex liquid state.

We thank J.M. Speight for discussions and S. Shevchenko for communications and
for informing us about a
perturbative study of the weak dipolar coupling regime  \cite{shev}
and a study of Josephson coupling  in DEC \cite{josephson}.

\end{document}